\documentclass[useAMS,usenatbib]{mn2e}
\usepackage{graphicx}
\usepackage{rotating}
\usepackage{lscape}
\usepackage{mdwtab}
\usepackage{amssymb}

\newcommand{\kms}{km~s\ensuremath{^{-1}}}
\newcommand{\cmdue}{cm$^{-2}$}
\newcommand{\cmtre}{cm$^{-3}$}
\newcommand{\um}{$\mu$m}
\newcommand{\hctren}{HC$_3$N}
\newcommand{\methanol}{CH$_3$OH}
\newcommand{\hdueco}{H$_2$CO}

\def \arcsec{\hbox{$^{\prime\prime}$}}
\def \arcmin{\hbox{$^{\prime}$}}

\title[The B1 shock in the L1157 outflow as seen at high spatial resolution]{The B1 shock in the L1157 outflow as seen at high spatial resolution\thanks{Based on observations carried out with the IRAM Plateau de Bure Interferometer. IRAM is supported by INSU/CNRS (France), MPG (Germany) and IGN (Spain).}}
\author[Benedettini et al.]{M. Benedettini$^{1}$\thanks{E-mail: milena.benedettini@inaf.it}, S. Viti$^{2}$, C. Codella$^{3}$, F. Gueth$^{4}$, A. I. G\'{o}mez-Ruiz$^{3}$,
\newauthor R. Bachiller$^{5}$, M.T. Beltr\'an$^{3}$, G. Busquet$^{1}$, C. Ceccarelli$^{6}$, B. Lefloch$^{6}$\\
$^{1}$INAF -- Istituto di Astrofisica e Planetologia Spaziali, via Fosso del Cavaliere 100, 00133 Roma, Italy \\
$^{2}$Department of Physics and Astronomy, University College London, Gower Street, London, WC1E6BT, UK\\
$^{3}$INAF -- Osservatorio Astrofisico di Arcetri, Largo E. Fermi 5, 50125 Firenze, Italy\\
$^4$Institut de Radio Astronomie Millim\'etrique, 300 Rue de la Piscine, F-38406, Saint Martin d'H\`eres, France\\
$^5$Observatorio Astron\'omico Nacional (IGN), Alfonso XII 3, 28014 Madrid, Spain\\
$^6$UJF-Grenoble 1/CNRS-INSU, Institut de Plan\'{e}tologie et d'Astrophysique de Grenoble UMR
5274, Grenoble, F-38041, France
}

\begin{document}

\date{Accepted . Received .}

\pagerange{\pageref{firstpage}--\pageref{lastpage}} \pubyear{2013}

\maketitle

\label{firstpage}

\begin{abstract}
We present high spatial resolution (750 AU at 250 pc) maps of the B1 shock in the blue lobe of the L1157 outflow in four lines: CS (3--2), CH$_3$OH (3$_{\rm K}$--2$_{\rm K}$), HC$_3$N (16--15) and p-H$_2$CO (2$_{02}$--3$_{01}$).
The combined analysis of the morphology and spectral profiles has shown that the highest velocity gas is confined in a few compact ($\approx$ 5\arcsec) bullets while the lowest velocity gas traces the wall of the gas cavity excavated by the shock expansion. 
A large velocity gradient model applied to the CS (3--2) and (2--1) lines provides an upper limit of 10$^6$ \cmtre\, to the averaged gas density in B1 and a range of 5$\times$10$^3\leqslant n_{\rm H_2} \leqslant$ 5$\times$10$^5$ \cmtre\, for the density of the high velocity bullets. The origin of the bullets is still uncertain: they could be the result of local instabilities produced by the interaction of the jet with the ambient medium or could be clump already present in the ambient medium that are excited and accelerated by the expanding outflow.
The column densities of the observed species can be reproduced qualitatively by the presence in B1 of a C--type shock and only models where the gas reaches temperatures of at least 4000 K can reproduce the observed HC$_3$N column density.

\end{abstract}


\section{Introduction}

\begin{table*}
  \caption{List of the observed transitions and observing parameters.}
  \begin{tabular}{@{}lccccc@{}}
  \hline
 transition & $\nu$ & E$_{up}$ & clean beam & resolution & rms \\
            & GHz   &  K   & \arcsec $\times$ \arcsec  & \kms  & mJy beam$^{-1}$ \\
 \hline
CH$_3$OH (3$_0$--2$_0$)E         & 145.094 & 27.1 & 3.48 $\times$ 2.31 (P.A.=12\degr) & 0.16 & 20 \\
CH$_3$OH (3$_{-1}$--2$_{-1}$)E2  & 145.097 & 19.5 & 3.48 $\times$ 2.31 (P.A.=12\degr) & 0.16 & 20 \\
CH$_3$OH (3$_0$--2$_0$)A$^+$     & 145.103 & 13.9 & 3.48 $\times$ 2.31 (P.A.=12\degr) & 0.16 & 20 \\
CH$_3$OH (3$_2$--2$_2$)E1, E2    & 145.126 & 36.2, 39.8 & 3.48 $\times$ 2.31 (P.A.=12\degr) & 0.16 & 20 \\
CH$_3$OH (3$_2$--2$_2$)A$^+$     & 145.133 & 35.0 & 3.48 $\times$ 2.31 (P.A.=12\degr) & 0.16 & 20 \\
HC$_3$N (16--15)                 & 145.561 & 59.4 & 3.47 $\times$ 2.30 (P.A.=12\degr) & 1.20 & 20 \\
p-H$_2$CO (2$_{02}$--1$_{01}$)   & 145.603 & 10.5 & 3.47 $\times$ 2.30 (P.A.=12\degr) & 1.20 & 20 \\
CS (3--2)                        & 146.969 & 14.1 & 3.43 $\times$ 2.25 (P.A.=10\degr) & 0.08 & 20 \\
\hline
\end{tabular}\\
\label{obs}
\end{table*}

Bipolar molecular outflows are one of the easily observable signatures of the early stages of the star formation process producing strong alterations of the protostellar environment both dynamically, accelerating the gas, and chemically, activating the high temperature chemistry in the shocked gas. One of the most interesting outflows is the one driven by L1157--mm, a low mass Class 0 protostar located at 250 pc \citep{looney07}. With respect to other outflows driven by low-mass protostars, the L1157 outflow stands out for its rich millimetre spectrum and it can be considered as the prototype of chemically active outflows. The outflow has been extensively observed mainly with single dish telescopes in many molecular lines such as CO (\citealt{umemoto92}; \citealt{bachiller01}), SiO (\citealt{zhang95}, 2000; \citealt{nisini07}), H$_2$ (\citealt{hodapp94}; \citealt{davis95}; \citealt{neufeld09}), NH$_3$ \citep{tafalla95}, and CH$_3$OH (\citealt{avery96}).
Two main shock events have been detected in the blue lobe of the outflow and the interferometric image of the CO (1--0) line \citep{gueth96} reveals that they are the apex of two cavities created by the propagation of large bow--shocks. The different orientation of the two cavities testifies the precession of the driving jet, even if a direct detection of the jet has not yet obtained. The brightest shock episode, called B1, is located at the apex of the second cavity and it has been extensively observed in several molecular species at millimetre (\citealt{gueth98}; \citealt{benedettini07}; \citealt{codella09}), near-infrared \citep{nisini10}, FIR and submillimeter wavelengths (\citealt{giannini01}; \citealt{nisini10b}; \citealt{lefloch10}, 2012; \citealt{codella10}; \citealt{benedettini12}), revealing a complex structure with various shock tracers peaking at different positions and the presence of gas at different excitation conditions.

Because of its chemical richness and its clear morphology, the L1157 outflow is an excellent laboratory to study the shock generated by protostellar outflows. Indeed it was observed by the Chemical Herschel Survey of Star forming regions (CHESS) $Herschel$ Key Program \citep{ceccarelli10} as prototype of chemically active outflows and it was used to test shocks models by many authors (e.g. \citealt{gusdorf08}; \citealt{flower10}, 2012). The results of this extensive modeling activity on L1157--B1 testify the complexity of this region where multiple shock types are acting. In fact different shock tracers have been explained with different types of shock: a pure C-type shock (\citealt{gusdorf08};  \citealt{neufeld09}; \citealt{viti11}), a dissociative J-type shock \citep{benedettini12} and a composition of the two called CJ-type shock (\citealt{gusdorf08b}; \citealt{flower10}, 2012). However, all the models have been applied to unresolved single dish data and hence the real structure of the shocked region could not be determined. We have now mapped L1157--B1 with the Plateau de Bure (PdB) interferometer in the 2 mm range, at a spatial resolution of about 3\arcsec\, corresponding to 750 AU at the distance of 250 pc for this source.
The observations are described in Sect. 2. We present the results in Sect. 3 and 4, the analysis in Sect. 5, 6 and 7 and the chemical modeling in Sect. 8. The main conclusions of the paper are summarized in Sect. 9.

\begin{figure}
\includegraphics[width=8cm,angle=0]{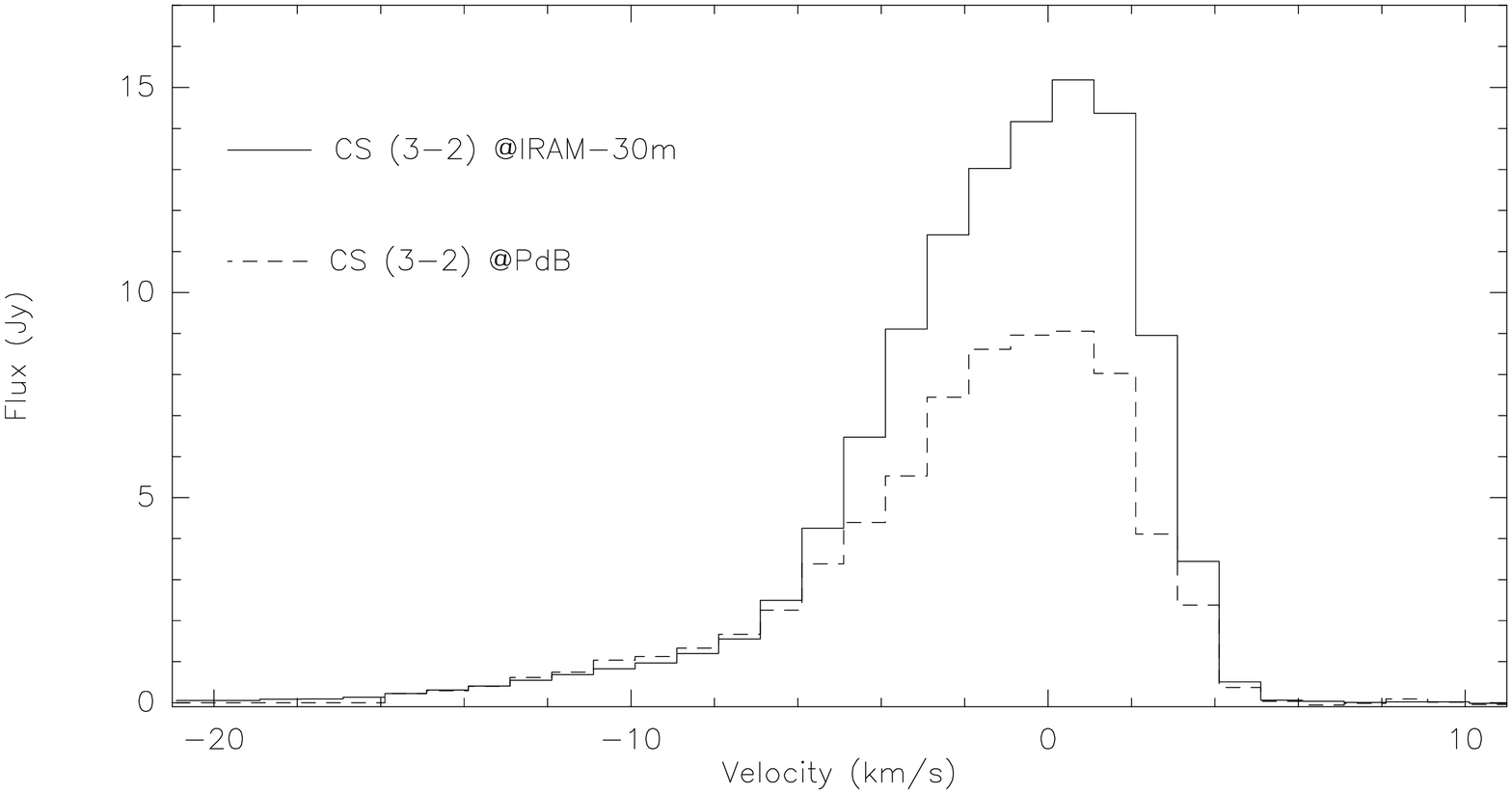}
\includegraphics[width=8cm,angle=0]{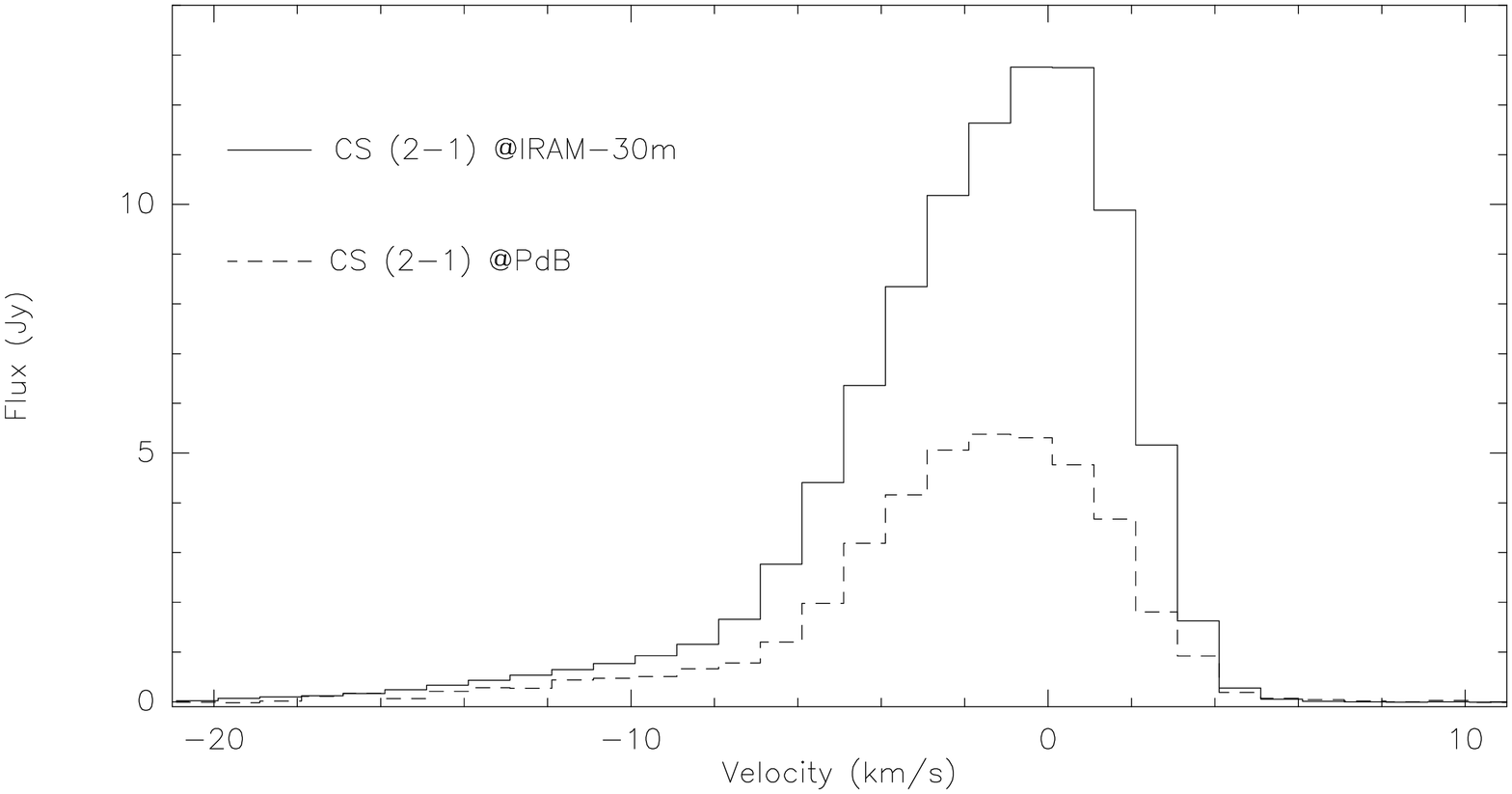}
\caption{{\it Top:} Comparison of the CS (3--2) spectrum acquired with the IRAM--30m (continuum line) \citep{gomez13} and the equivalent spectrum extracted from the PdB data (dashed line). {\it Bottom:} as above but for the CS (2--1) line.}
\label{iram+pdb}
\end{figure}

\section{Observations}

\begin{figure*}
\includegraphics[width=16cm,angle=0]{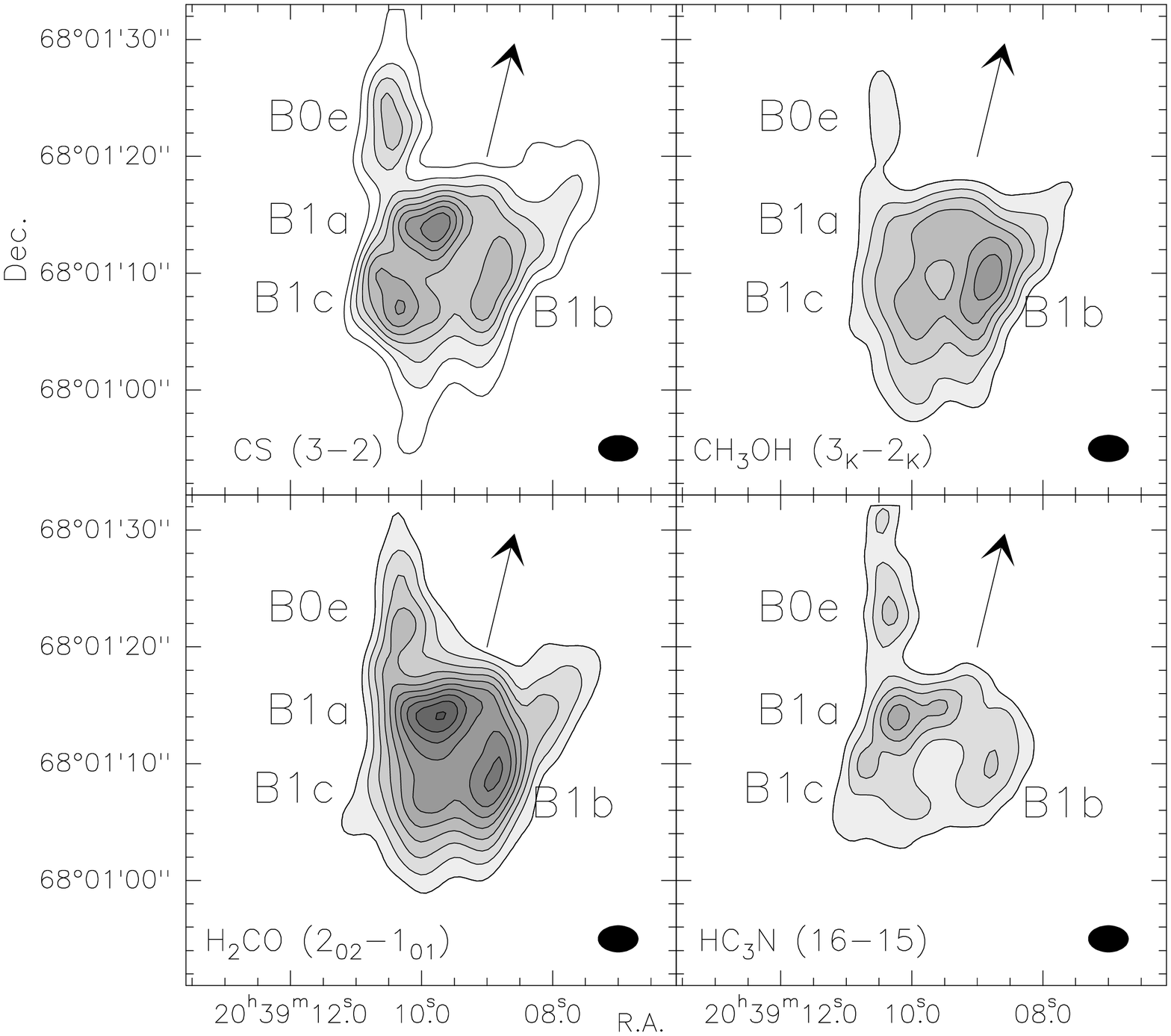}
\caption{Interferometric images of L1157--B1. The first contour and steps correspond to a 3$\sigma$ level and are: 0.75 Jy beam$^{-1}$ \kms\,for CS (3--2), 3 Jy beam$^{-1}$ \kms\, for CH$_3$OH (3$_{\rm K}$--2$_{\rm K}$), 0.54 Jy beam$^{-1}$ \kms\, for p-H$_2$CO (2$_{02}$--1$_{01}$) and 0.15 Jy beam$^{-1}$ \kms\, for HC$_3$N (16--15). The filled ellipse in the bottom left corner represent the HPBW, the arrow indicates the direction toward the central driving source L1157--mm.}
\label{sum_map}
\end{figure*}
The observations were carried out with the PdB interferometer between August 2008 and March 2009. The CD configuration was used with three configurations of 5 or 6 antennas. We simultaneously observed the CS (3--2) transition at 146.969 GHz in the Lower Side Band (LSB) and HC$_3$N (16--15), p-H$_2$CO (2$_{02}$--1$_{01}$) and CH$_3$OH (3$_K$--2$_K$), with frequency between 145.094 and 145.603 GHz in the Upper Side Band (USB). The 20, 40 and 160 MHz correlator units were used for covering different part of the USB and LSB, resulting in a spectral resolution ranging from 0.08 to 1.2 \kms. The phase and amplitude calibration was achieved by observations of 1849+670 and 2021+614. The bandpass of the receivers were calibrated by observations of 3C84 and 3C454.3. The flux calibration was determined relative to 3C84 and 3C454.3, with an uncertainty of ∼20\%. 
The data were calibrated and analyzed with the GILDAS\footnote{http://www.iram.fr/IRAMFR/GILDAS} software. Images were produced using natural weighting and were restored with a final clean beam of 3.48\arcsec$\times$2.31\arcsec\, at 145 GHz.
The details of the observations are summarized in Table \ref{obs}. In particular, the emission at 145.126 GHz is the contribution of two methanol lines (3$_ 2$--2$_2$)E1 and (3$_ 2$--2$_2$)E2 very close in frequency (145.126191 and 145.126386 GHz) with very similar Einstein coefficients (6.77$\times$10$^{-6}$ s$^{-1}$ and 6.86$\times$10$^{-6}$ s$^{-1}$) and upper level energies (36.2 and 39.8 K), therefore the flux emitted in the two lines should be very similar. We attributed the flux measured at this frequency half to each of the two lines.

\section{Estimate of the filtering of the large scale emission}
\label{filtering}

With interferometers the largest sensitive structure depends on the shortest baseline which is 14 m for our observations. Hence our 2 mm observations can only correctly measure the flux of structures smaller than $\sim$13\arcsec\, as, for larger structures, part of the flux is missed out. 
In order to evaluate the missing flux in the lines observed with PdB we compared the spectrum of each transition produced summing the emission measured at PdB in a circle of diameter equal to the HPBW of the IRAM--30m telescope, with the spectrum measured at the single dish IRAM-30m \citep{gomez13}. In Fig. \ref{iram+pdb} we show the two spectra of the CS (3--2) line. We found that the percentage of the missing flux depends on the velocity: for the high velocity gas ($v<$ - 6 \kms) PdB recovers 100\% of the flux while for the low velocity gas ($v>$ - 6 \kms) the flux measured at PdB is $\sim$ 62$\%$ of the IRAM flux. Similar percentage of missing flux is also found for the other observed lines HC$_3$N (16--15) and p-H$_2$CO (2$_{02}$--1$_{01}$) while for methanol is not possible to evaluate the missing flux because the lines are blended. We performed the same analysis for the CS (2--1) line previously observed with PdB \citep{benedettini07} with
the same CD configuration but a slightly higher value of the shortest baseline (19 m), resulting in a similar value of the largest sensible structure (14\arcsec) and spatial resolution (3.11\arcsec$\times$ 2.79\arcsec) of the 2 mm observations. Unlike the case of the CS (3--2) transition, for the CS (2--1) line the flux filtering affects all the velocities (see Fig. \ref{iram+pdb}). In particular, for the high velocity gas ($v<$ - 6 \kms) about half ($\sim$ 50$\%$) of the IRAM flux in the (2--1) line is missed in the PdB spectrum while for $v>$ -6 \kms\, a similar percentage ($\sim$ 57$\%$) of filtering is measured in the two lines.
However, the largest sensitive structure for the CS (2--1) line is similar to the value of the CS (3--2), therefore the difference in the missing flux indicates that a consistent part of the emission of the CS (2--1) line comes from extended ($>$14\arcsec) structures also for high velocity gas and therefore is filtered in the PdB spectrum while the emission of this high velocity structure in the CS (3--2) is negligible (see Sect. \ref{g2} for further discussion).

\section{Results}

Line maps of the four observed transitions are shown in Fig. \ref{sum_map}. The morphology is well in agreement with that of other molecules (e.g. \citealt{gueth98}; \citealt{benedettini07}) confirming the clumpy structure superimposed to the more extended arch--like shape seen e.g. in CO \citep{gueth96}. It is worth noting that the 
spatial resolution of these 2 mm maps ($\sim$ 3\arcsec) is the best compromise for detecting both the extended and the compact gas.
The clumps B0e, B1a, B1b and B1c, already identified by \cite{benedettini07} with previous PdB observations at 3 mm, are clearly present also in these higher energy transitions. It is also worth noting that while the CS and \methanol\, emission have a morphology very similar to that observed in lower excitation lines (with \methanol\, brighter in the west-side clumps (B1b) and CS brighter in the east-side clumps (B1a, B1c)), \hctren\, and \hdueco\, are brighter in the north-side clumps (B1a), similar to what is observed for CH$_3$CN (8$_K$--7$_K$) \citep{codella09}. These results confirm the complexity, both in the morphology and in the chemistry, of the B1 region, likely induced by the shock originated by the interaction between the driving precessing jet and the ambient material.

\begin{figure*}
\includegraphics[width=17cm]{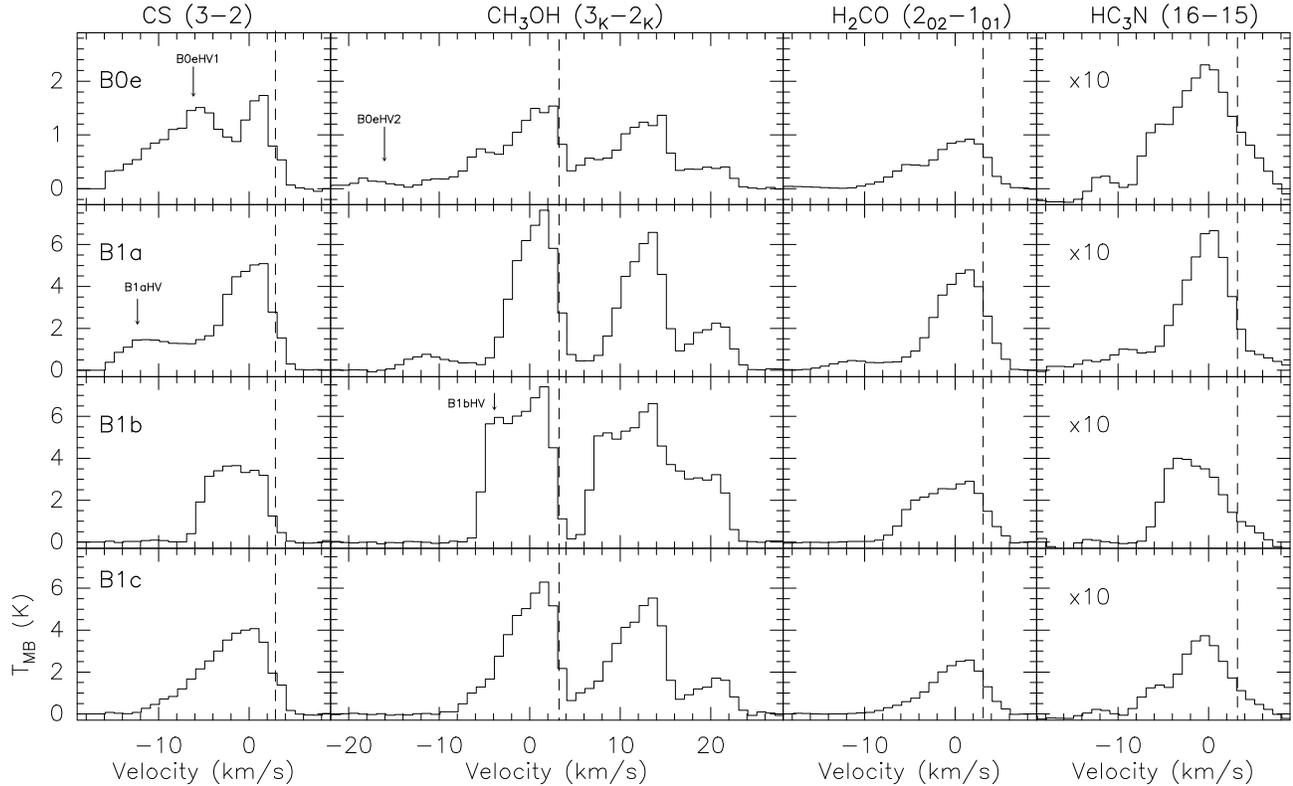}
\caption{Spectra of the observed lines in the four clumps. The spectra are resampled at a common resolution of 1 \kms. The dashed line indicates the systemic velocity $v_{\rm lsr}$=2.6 \kms. Arrows indicate the peak velocity of the HV bullets. The CH$_3$OH (3$_{\rm K}$--2$_{\rm K}$) spectra show three methanol lines and the reference frequency is 145.103185 GHz.}
\label{spettri}
\end{figure*}

\begin{figure*}
\includegraphics[width=16cm,angle=0]{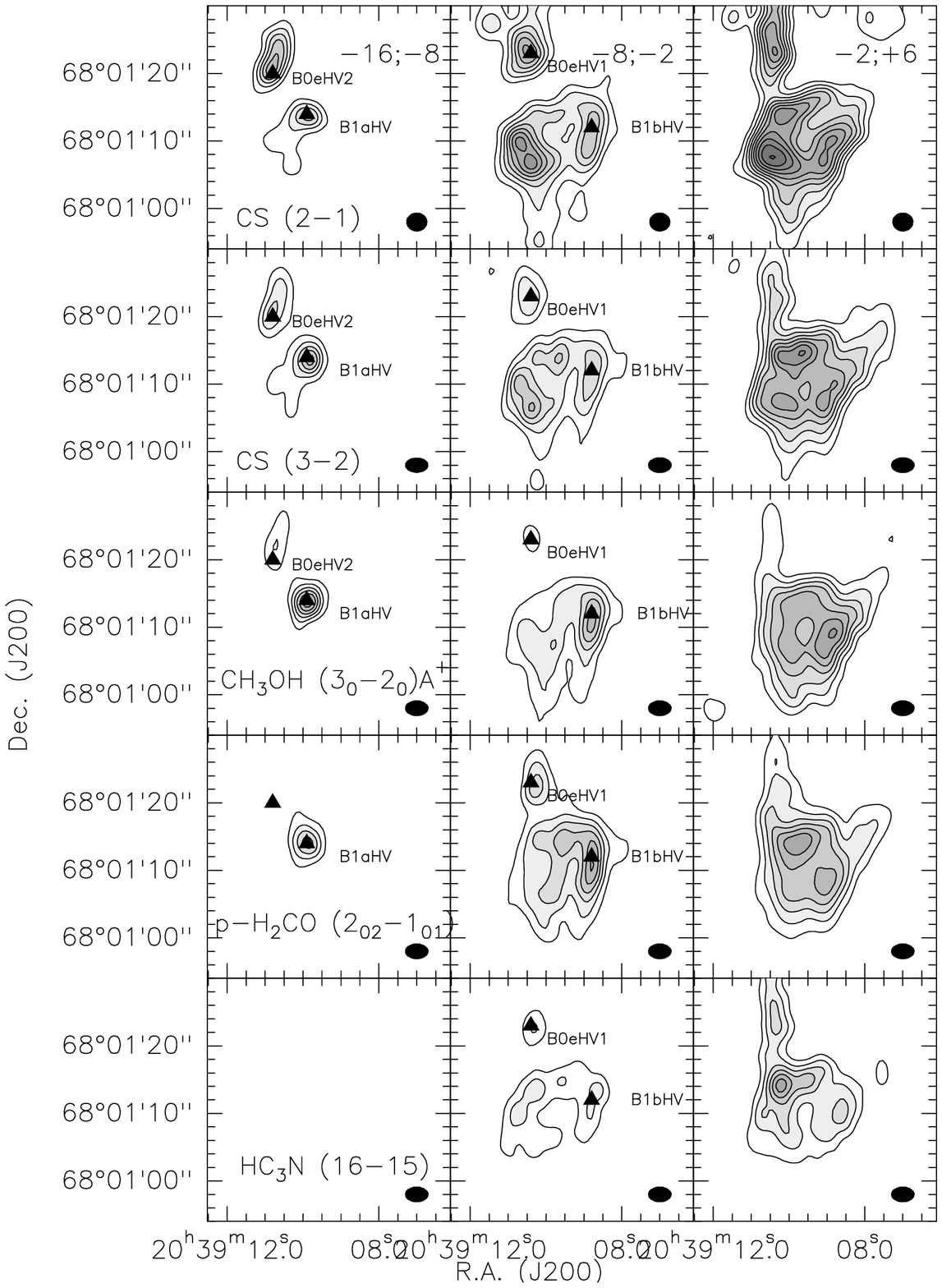}
\caption{Channel maps of the four observed transitions plus the CS (2--1) \citep{benedettini07}. The first contour and contour steps correspond to a 3$\sigma$ level. Triangles mark the position of the HV bullets. In the left panels the velocity limits are --16$<v<$--8 \kms\, and all the flux is recovered while in the central panels (--8$<v<$--2 \kms) and in the right panels (--2$<v<$+6 \kms) about 60\% of the extended emission is filtered out.}
\label{channel}
\end{figure*}

Differences along B1 are also present in the profiles of the observed lines (see Fig. \ref{spettri}). In most of the observed area the line emission ends at $\sim$+6 \kms\, in the red wing and at $\sim$-10 \kms\, in the blue wing. However, toward B0e, B1a and B1b the blue emission extends at velocities lower than -10 \kms\, and the spectra show multiple peaks indicating the presence of different gas components toward these lines of sight. We labeled these spectrally identified additional components flowing at higher velocity with the HV suffix in order to distinguish them from the gas component flowing at velocity closer to the systemic velocity (2.6 \kms).
We found four high velocity (HV) bullets:
\begin{itemize}
 \item in B1a a second peak (called B1a--HV) is detected at $v_{\rm lsr}$= --12 \kms, the secondary peak is visible in the spectra of all the observed species;
 \item in B0e a second peak (called B0e--HV1) is detected at $v_{\rm lsr}$= --6 \kms, the secondary peak is visible in the spectra of all the observed species;
 \item in B0e a third peak (called B0e--HV2) is detected at $v_{\rm lsr}$= --16 \kms; this peak is only partially visible in the CS (3--2) spectrum because no data were taken for velocities $<$--16 \kms\, but it is visible in the CH$_3$OH spectrum and also in the CS (2--1) \citep{benedettini07};
 \item in B1b a second peak (called B1b--HV) is detected at $v_{\rm lsr}$= --4 \kms, the intensity of the B1b--HV secondary component is similar to the ambient component in all the lines but HC$_3$N (16--15) where it clearly dominates the emission.
\end{itemize}

The morphology of B1 at various velocities is shown in the channel maps of Fig. \ref{channel}. The emission at the line centre (right panels of Fig. \ref{channel}) clearly show the walls of the CO cavity and the peaks of the brightest clumps. The emission at intermediate velocities (central panels of Fig. \ref{channel}) shows still some contribution from the cavity but some clumps emerge (B0e--HV1 at $v_{\rm lsr}$= --6 \kms\, and B1b--HV at $v_{\rm lsr}$= --4 \kms). Finally, the emission at the highest velocities is well confined in only two small clumps B1a--HV at $v_{\rm lsr}$= --12 \kms\, and B0e--HV2 at $v_{\rm lsr}$= --16 \kms. 
In Table \ref{clumps_coor} the position, the peak velocity and the deconvolved size (from 3\arcsec\, to 6\arcsec) of the identified clumps are listed. 

Clumpy emission at high velocity, the so called extreme high velocity (EHV) bullets, have been already observed in some outflows driven by low mass protostars, as for example L1148-mm, IRAS 04166+2706 and HH211 (\citealt{dutrey97}; \citealt{nisini07}; \citealt{gueth99}; \citealt{santiago09}), in the form of a sequence of bullets aligned along the jet that are thought to be associated to episodic ejection events in the jet. Some substantial differences induce us to assert that the HV bullets observed in L1157-B1 are different in nature with respect the EHV bullets. In fact, 
i) our bullets have less extreme velocities (from 7 to 19 \kms\, with respect to the $v_{\rm lsr}$ of the cloud) than the EHV bullets which have velocity up to 50 \kms;
ii) they are not aligned on a single line along the jet but they have an arch shape and are associated with the walls of the cavity excavated by the jet; 
iii) they are chemically rich being observed in several molecular species while the EHV bullets have only be seen in CO and SiO.
Therefore the L1157-B1 bullets are probably related to local instabilities of the low excited material swept up by the outflow wind rather than to the highly excited material of the jet (see Sect. \ref{bo} for further discussion).

\begin{table}
\caption{List of the HV bullets.}
\begin{tabular}{@{}lccrc@{}}
 \hline
clump & R.A.(J2000) & Dec(J2000) & $v{\rm (peak)}$ & size\\
    & h~ m ~s ~ & $\degr$ ~ $\arcmin$~~ $\arcsec$& \kms & $\arcsec$\\
 \hline
B0e--HV2 & 20 39 10.8 & 68 01 20 & --16 & 3\\
B0e--HV1 & 20 39 10.6 & 68 01 23 &  --6 & 5\\
B1a--HV  & 20 39 09.9 & 68 01 14 & --12 & 4 \\
B1b--HV  & 20 39 08.8 & 68 01 12 &  --4 & 6 \\
 \hline
\end{tabular}
\label{clumps_coor}
\end{table}

\section{Analysis of the CS lines}

\subsection{Lines profiles}
\label{g2}

Analyzing the profile of the CO lines with $J_{up}\le$ 16 observed in L1157--B1 with $Herschel-HIFI$, \cite{lefloch12} found a common line shape of the CO lines that has been described as a combination of three exponential laws, $I(v)\propto$ exp (-$|v/v_0|$), that, based on their temperature, the authors associated to different gas components of the outflow. In particular, the intermediate excitation component, called $g2$, with $n_{\rm H_2}\ge$ 10$^5$ \cmtre\, and $T$ = 64 K, has been associated to the gas of the walls of the outflow cavity produced by the B1 shock, while the low excitation component, called $g3$, with $n_{\rm H_2}\ge$ 10$^5$ \cmtre\, and $T$ = 23 K, has been associated to the gas belonging to the second cavity of the blue lobe produced by a shock event -- B2 -- older than B1.

\begin{figure}
\includegraphics[width=6.5cm,angle=90]{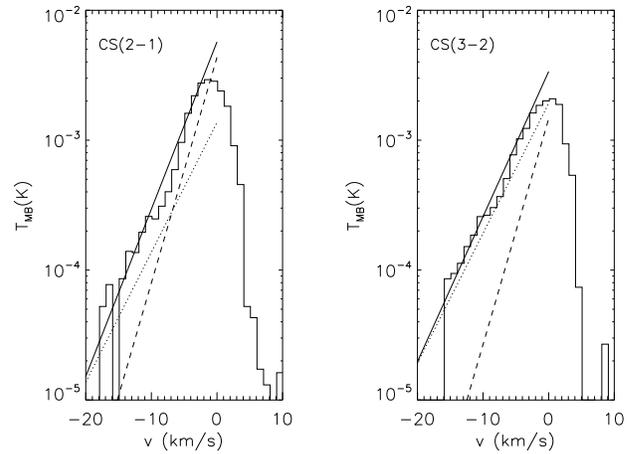}
\caption{Spectra of the CS (2--1) and CS(3--2) lines extracted from the PdB maps convolved with a beam of 37$\arcsec$ (the HIFI beam at 576 \micron). The spectra are compared with the exponential profiles of the $g2$ (dotted line) and $g3$ (dashed line) components defined by \citet{lefloch12}. The continuum line is the linear composition of the two exponential fits.}
\label{HIFI_g2g3}
\end{figure}

We searched for the same spectral signatures in the CS lines observed with PdB. For the comparison of our data with the $Herschel-HIFI$ spectra, we convolved the CS PdB maps to the $HIFI$ spatial resolution at the frequency of CO (5--4) line, i.e. 37\arcsec, and extracted the spectra over the same beam. Both the CS (2--1) and the CS (3--2) line profiles can be well fitted by a composition of the two exponential laws $g2$ and $g3$ with the same parameters found for the CO lines (see Fig. \ref{HIFI_g2g3}). This indicates that the CS emission at large spatial scale, that dominates the line profile in the maps convolved at higher spatial resolution, arises from the same gas component traced by low and intermediate $J_{up}$ CO lines.
Our PdB maps show that this large scale emission originates from the walls of the B1 cavity therefore confirm the association of $g2$ with the wall of the B1 cavity as suggested by \cite{lefloch12}. 
A posteriori we verified that the flux filtering of this extended gas component produced by the interferometer did not modify significantly the line shape. Indeed the spectral profiles of the two CS (3-2) and (2-1) lines observed with the single dish IRAM - 30m \citep{gomez13} are well fitted by the same two exponential laws that fit the PdB spectra convolved at 37" resolution.

From Fig. \ref{HIFI_g2g3} we see that for the CS (2--1) line $g2$ and $g3$ give a similar contribution to the total flux while the CS (3--2) line is dominated by $g2$, testifying the higher excitation condition of $g2$ with respect to $g3$. Moreover, $g3$ is also associated with a gas at a velocity lower than $g2$, this is a further indication of its association with the slower gas of the older B2 cavity \citep{benedettini07}. The gas belonging to the B2 cavity intercepted by the line of sight toward B1 is certainly extended and therefore suffers of filtering in our interferometric observations. In order to quantify this evidence, we calculated the flux of each $g2$ and $g3$ component integrating the two exponential laws in the same velocity range. We found that for $v<$ --6 \kms\, the $g3$ component in CS (3--2) is the 14\% of the total flux while in CS (2--1) is the 40\%. This result explains the different percentage of the filtering at $v<$ --6 \kms\, observed in the two CS lines and discussed in Sect. \ref{filtering}.

\subsection{LVG modeling}
\label{lvgfit}

The physical conditions of the gas component in outflows can be derived by solving the radiative transfer problem in the line simultaneously with the level populations, under the Large Velocity Gradient (LVG) approximation. In this approximation the line flux depends on the escape probability that is a function of the line optical depth, which in turn is proportional to the number density of the considered molecular species n(x) and inversely proportional to the velocity gradient dV/dz. Assuming an homogeneous slab where the total velocity dispersion is V, the factor n(x)$\times$dz/dV can be expressed in term of N(x)/V where N(x) is the column density of the particular molecule, making explicit the dependency of the line flux from the column density. We used the LVG code developed by \citet{ceccarelli03} in a plane parallel geometry using the molecular parameters found in the BASECOL\footnote{http://basecol.obspm.fr} database \citep{dubernet06}. In particular, we modeled the first 31 levels of CS using the collisional coefficients with H$_2$ from \citet{turner92}. We adopted a line width of 10 \kms, as usually done from previous models of this region. 

\begin{figure}
\includegraphics[width=8cm]{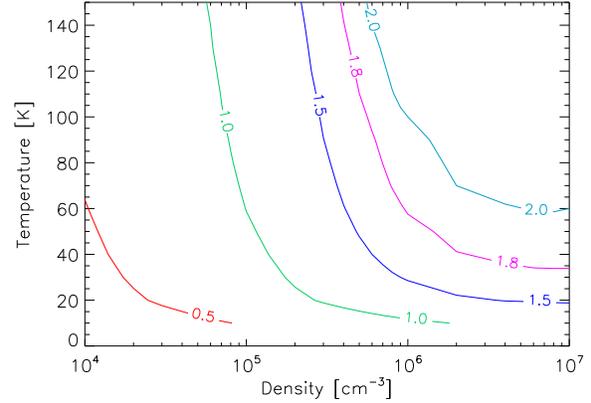}
\caption{Theoretical CS (3--2)/(2--1) line ratio as function of density and temperature for a line width of 10 \kms and a CS column density of 3$\times$10$^{13}$ \cmdue, i.e. the typical value derived in B1 (see Table \ref{column_bullets}). The behavior of the ratio for column densities in the range 10$^{13}$ -- 10$^{15}$ \cmdue\, is very similar.}
\label{cstheoratio}
\end{figure}

Because of the filtering affecting the extended emission, a quantitative LVG analysis can be carried out for the compact HV gas but not for the diffuse gas component. Nonetheless, we can use the observed CS (3--2)/(2--1) line ratio to give some general constraints on the physical conditions of the gas. In fact, as shown in Sect. \ref{filtering}, the filtering affects the (2--1) line more than the (3--2) so that the observed line ratio (CS (3--2)/(2--1) = 1.8) can be considered as an upper limit. In Fig. \ref{cstheoratio} we show the theoretical CS (3--2)/(2--1) line ratio calculated under the LVG approximation. For temperatures higher than $\sim$ 60 K the ratio is quite insensitive to temperature changes and it mainly depends on the gas density. However constraints on the gas temperature can be derived from other tracers. In particular, \citet{codella09} measured the temperature in the different B1 clumps by using a rotational diagram of CH$_3$CN lines observed with PdB in the same configuration of the data of this paper and they found temperatures ranging from 55 to 132 K. This range is compatible with other temperature estimates, namely 80 K derived by \cite{tafalla95} by means of VLA observations of NH$_3$ and 64 K derived by \citet{lefloch12} from CO line observed with $Herschel - HIFI$. For our analysis of L1157--B1 we assume a temperature ranging from 55 to 132 K. Under this constrain the upper limit of 1.8 for the CS (3--2)/(2--1) ratio implies an upper limit on the averaged gas density in B1 of 10$^6$ \cmtre. This limit refines previous estimates of the gas density in B1 that give only a lower limit of $\geqslant$ 10$^5$ \cmtre. In particular, the CO lines from {\it Herschel} indicate $n_{\rm H_2}\geqslant$ 10$^5$ \cmtre, for the higher excited ($T>$ 200 K) gas (\citealt{benedettini12}; \citealt{lefloch12}). SiO lines indicate $n_{\rm H_2}\geqslant$ 3$\times$10$^5$ \cmtre\, and $T$=150--300 K \citep{nisini07}. Note however that these lower limits have been derived only from single dish observations and in some cases from lines with higher excitation temperatures with respect to the ones used in this study.

\begin{figure}
\includegraphics[width=8.5cm]{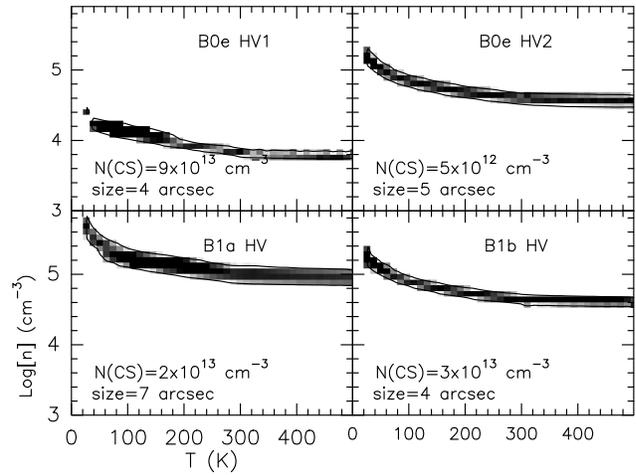}
\caption{$\chi^2$ distribution in the temperature -- density plane for the LVG fitting of the CS lines in the four HV bullets. The black area shows the parameters space with a reduced $\chi^2\le$1 for the CS column density and size written in the panels.}
\label{chisquare}
\end{figure}

\begin{table}
\caption{Results of the LVG fitting in the HV bullets.}
\begin{tabular}{@{}lccc@{}}
 \hline
clump & $N$(CS) & size    & $n_{\rm H_2}$ \\
      & \cmdue    & $\arcsec$& 10$^4$ \cmtre \\
 \hline
B0e--HV2 & 4(12)-1(13) & 2-8 & 5-10\\
B0e--HV1 & 6(13)-1(14) & 4-8 & 0.5-1\\
B1a--HV  & 2(13)-7(13) & 2-8 & 5-50 \\
B1b--HV  & 2(13)-8(13) & 2-8 & 2-10 \\
 \hline
\end{tabular}
\label{lvg}
\end{table}

The compact emission from the HV bullets having size $\lesssim$ 6\arcsec, lower than the largest sensitive structure ($\sim$14\arcsec) of the interferometer does not suffer of flux filtering therefore we did a quantitative analysis of this emission. In particular, we used the LVG code to derive the physical conditions of the four HV bullets B0e--HV1, B0e--HV2, B1a--HV and B1b--HV by using the two CS lines (3--2) and (2--1). In Fig. \ref{channel} we show that the HV bullets are detected in both lines. 
To measure the flux emitted from the HV bullets we integrated the line emission over the velocity range associated to the specific spectral component. In this way we are also considering the (minor) contribution of the extended emission that can be present, especially in the spectral component closer to ambient velocity. However, it is impossible to deblend the flux of the HV bullet from the flux of the more extended component because we do not know the shape of the line profile of the accelerated gas that is not Gaussian. To take into account this additional uncertainty, we associated an error of 20\% to the derived line fluxes. In Fig. \ref{chisquare} we show the distribution of the reduced $\chi^2$ as function of temperature and density for best fit values of size and column density. 
It is well known that CS is a good indicator of the gas density: by using the two CS lines and assuming the size measured in the maps (5\arcsec$\pm$3\arcsec) we can constrain quite well the CS column density and the number density of the gas. On the contrary, the temperature remains undefined over a wide range, from a few tens to a few hundreds Kelvin.
In Table \ref{lvg} we report the results of the LVG fitting giving the range of the CS column density and gas density that fit the data under the adopted clump size of (5\arcsec$\pm$3\arcsec) and gas temperature (55 -- 132 K). Note that the flux of the B0e--HV2 clump is underestimated because the CS (3--2) spectrum does not cover the whole spectral range of the line, as a consequence the results are approximated and in particular the column density is underestimated. The results of the LVG fitting indicate a lower gas density ($n_{\rm H_2}\leqslant$ 10$^4$ \cmtre) for the B0e--HV1 bullet with respect to the other three bullets ($n_{\rm H_2}\geqslant$ 5$\times$10$^4$ \cmtre). In general the gas of the HV bullets (5$\times$10$^3\leqslant n_{\rm H_2} \leqslant$ 5$\times$10$^5$ \cmtre) seems to be less dense than the large scale emitting gas (10$^5 \leqslant n_{\rm H_2} \leqslant$ 10$^6$ \cmtre).

\section{CH$_3$OH rotational diagram}

A rough estimate of the gas temperature can be derived by using the rotational diagram. If the emission lines are optically thin the slope of the diagram gives directly 1/$T_{\rm rot}$ and the rotational temperatures is a lower limit to the kinetic temperature if the gas is not in Local Thermodynamic Equilibrium (LTE). 

\begin{figure}
\includegraphics[width=6cm,angle=90]{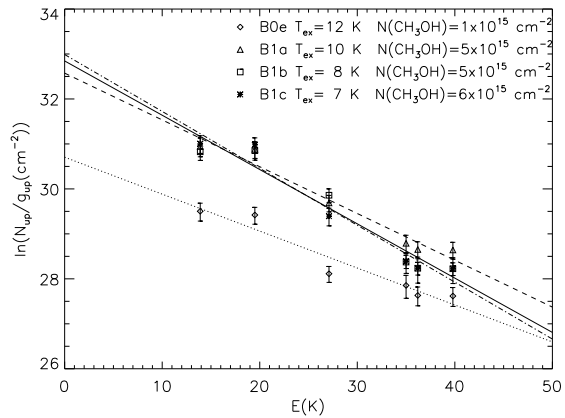}
\caption{CH$_3$OH rotational diagrams with superimposed linear fits. Diamonds, triangles, squares and stars correspond to clump B0e, B1a, B1b and B1c, respectively. The inferred excitation temperatures and total CH$_3$OH column densities for each clump are reported in the figure.}
\label{ch3oh_rotdiagram}
\end{figure}

Since the observed methanol transitions are quite close in frequency in most clumps the lines are blended (see Fig.\ref{spettri}) and it is not possible to evaluate the contribution of the various spectral components, i.e. the HV clumps. In those cases where it is not possible to deblend the contribution of the HV and LV components we attributed all the flux at a certain velocity to the dominant component. Out of the four HV bullets only for B1b--HV enough lines can be measured to build the rotational diagram, and we find $T_{\rm{rot}}$ = 14~K and $N$(CH$_3$OH) = 3$\times$10$^{15}$~\cmdue. We also built the rotational diagrams of the five methanol lines for the low velocity gas in each clump (see Fig.~\ref{ch3oh_rotdiagram}). We find a similar rotational temperature $T_{\rm{rot}}\simeq10$~K at all positions, indicating that the excitation conditions are rather uniform in B1. The total column density of CH$_3$OH is $\simeq$ 5$\times$10$^{15}$~\cmdue\ apart in B0e where it is slightly lower 1$\times$10$^{15}$~\cmdue. The derived rotational temperature is in agreement with previous estimates based on single dish CH$_3$OH \citep{bachiller95} and $^{13}$CH$_3$OH \citep{codella12} data. This quite low temperature indicates that the methanol molecule is subthermally excited. Therefore $T_{\rm rot}\ll T_{\rm kin}$ and the temperature derived from the rotational diagram cannot be considered as an estimate of the gas kinetic temperature.
On the other hand, in the subthermal regime the critical density of the lines gives an upper limit to the gas density. The critical density of the observed methanol lines is of the order of 10$^5$ \cmtre, compatible with the densities of the HV bullets that we found in the previous section from the LVG analysis of CS.

\section{Zooming in B0e}
\label{bo}

\begin{figure}
\includegraphics[width=7cm]{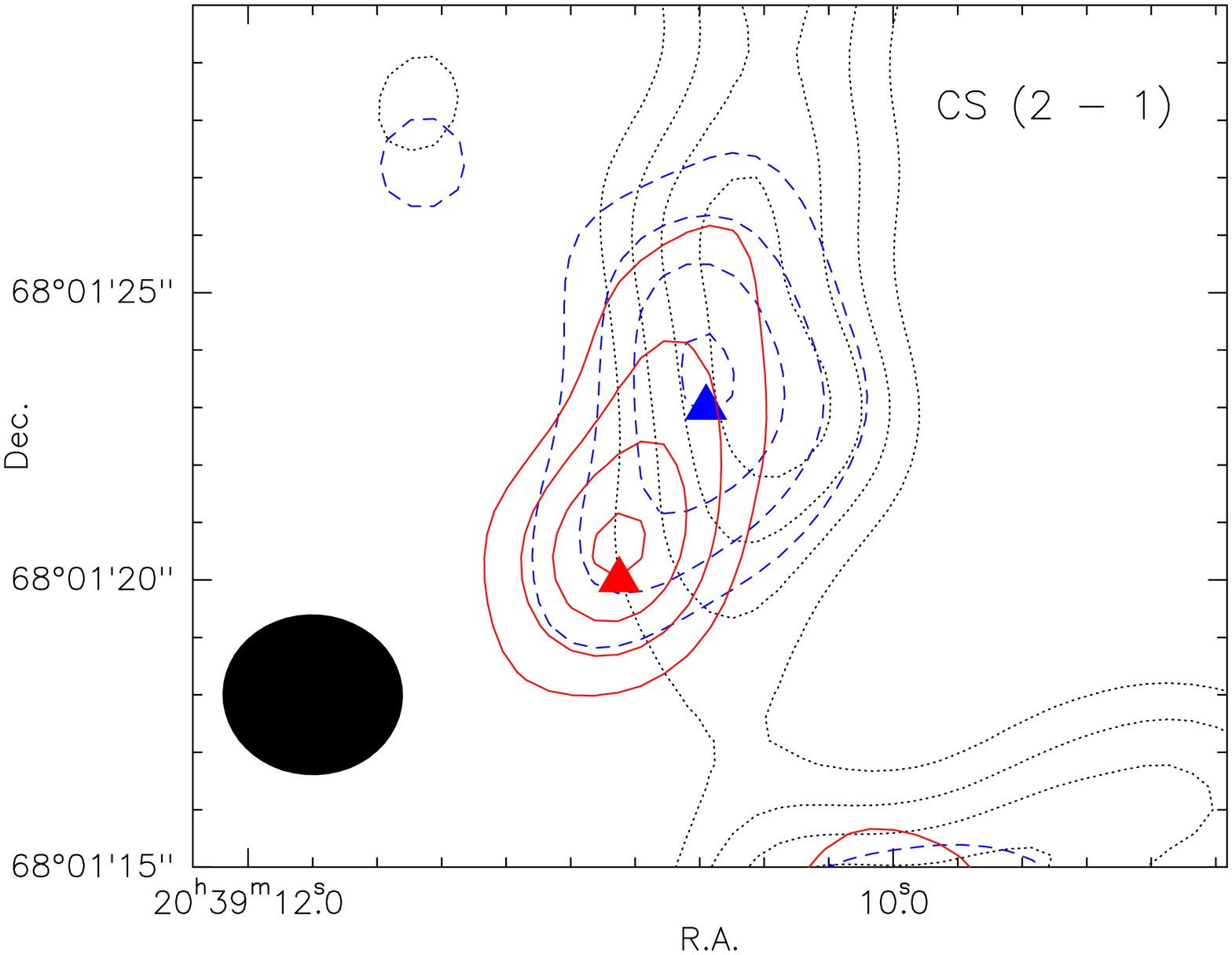}
\includegraphics[width=7cm]{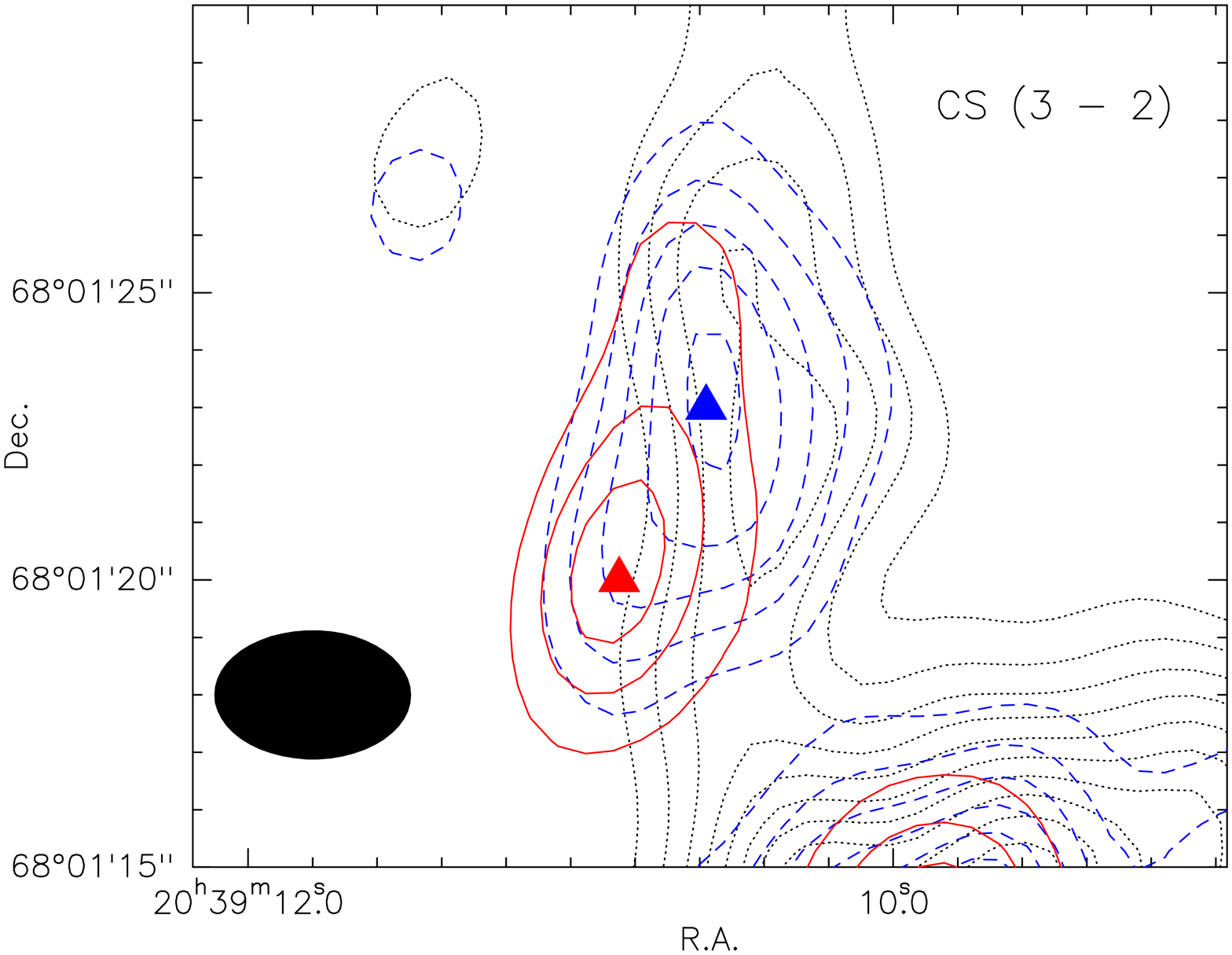}
\caption{The B0e clump. {\it Top:} CS (2--1) line intensity integrated over three velocity ranges:  -20.5,--13 \kms\, (red solidline) level steps 0.1 Jy beam$^{-1}$ \kms, --13,--3 \kms\, (blue dashed line) first level 0.5 Jy beam$^{-1}$ \kms\, steps 0.3 Jy beam$^{-1}$ \kms, --3, +5 \kms\, (black dotted line) first level 0.5 Jy beam$^{-1}$ \kms\, steps 0.3 Jy beam$^{-1}$ \kms. The triangles mark the peak of the emission in each velocity range. The filled ellipse shows the HPBW of the map. {\it Bottom:} the same for CS (3--2).}
\label{csb0}
\end{figure}

\begin{figure}
\includegraphics[width=8cm]{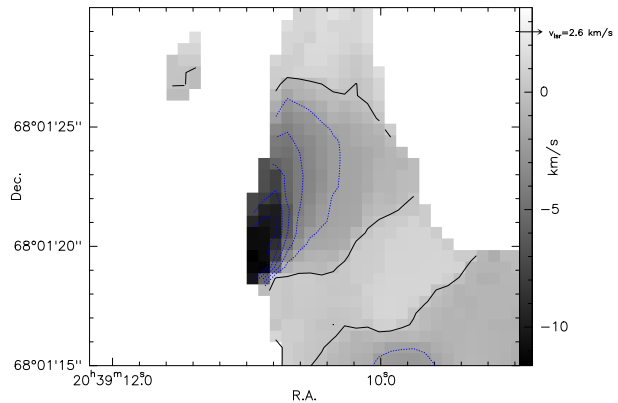}
\caption{CS (3--2) first moment toward B0e. Continuum line indicates the 0 velocity and dotted lines the negative velocities.}
\label{cs_moment}
\end{figure}

In Fig. \ref{csb0} we plotted the CS (3--2) and (2--1) emission from the three spectral components detected toward the B0e region, i.e. the two HV bullets B0e--HV2 at --16 \kms\, and B0e--HV1 at --6 \kms\, and the low velocity gas peaking at 1.2 \kms. The morphology of the three components is very similar in the two lines. As one can see the position of the peaks of the two highest velocity clumps B0e--HV1 and B0e--HV2, is external with respect to the walls of the cavity traced by the low velocity gas. The distance between the peak position of B0e--HV1 and the peak of the LV gas in B0e is 1\farcs4, i.e. 1/3 of the spatial resolution of the map, so it may not be significant even if the signal to noise ratio of the map is quite good. On the other hand, the distance of the B0e--HV2 with respect to  B0e--HV1 and B0e LV is 3\farcs2 and 3\farcs0, respectively, i.e. of the same order of the HPBW, hence it should be a real feature indicating that the compact bullet with the highest velocity is at different position of the slower gas and in particular it is external to the walls of the outflow cavity.
In addition, the first moment of both CS lines (Fig. \ref{cs_moment}) shows a gradient with mean velocity  increasing from the internal wall of the cavity toward the outside in the direction of B0e--HV2.

The interpretation of these very intriguing features is not univocal.
It could be the precessing jet that is now impacting at the B0e position producing an acceleration of the gas outwards of the cavity. Presently there is not any direct detection of the jet driving the L1157 outflow. On the other hand, a large set of shock tracers such as high velocity SiO (2--1) \citep{gueth98}, [FeII ]at 26 \um\, \citep{neufeld09}, H$_2$ $v=$1-0\,S(1) \citep{caratti06}, [OI], OH and high--J$_{up}$ CO \citep{benedettini12} all peak toward B1a indicating that this is the point with the highest excitation conditions and therefore suggesting that it should be the position where the jet is impacting the lower velocity gas of the outflow. However, the higher gas velocities as well as the velocity gradient in B0e (Fig. \ref{cs_moment}) seem to imply that the jet is moving toward the B0 position as also foreseen by the model of the jet precession by \citet{gueth98}. In this scenario the observed clumpiness could be the result of local instabilities produced by the interaction of the driving jet and the ambient medium. Alternatively, the HV bullets could be clumps already present in the ambient medium before the advent of the outflow that are compressed and pushed by the expanding outflow cavity. The clumpiness of the interstellar medium is well known (e.g. \citealt{viti03}, \citealt{morata05}) and the possible pre--existence of the clumps observed in outflows has been proposed by some authors (\citealt{viti04a}; \citealt{benedettini06}). The slightly lower density that we found in the compact bullets with respect to the bulk of the outflowing gas can support the hypothesis that the clumps are not formed by the outflow but can, at least partially, be present in the cloud before the arrival of the shock and pushed by the shock front. Indeed, from a rough estimate, the momentum of the B0e--HV2 bullet is similar to the momentum of the cavity as expected from the law of the conservation of momentum in collision. 

\section{The chemistry}

\begin{table*}
\caption{Column density of the observed species calculated in each clump for two different temperatures. The notation used is a(b) that means a$\times$10$^b$.}
\begin{tabular}{llrcc}
\hline
 clump & transition &  Flux  & $N$(X$_i$) ($T$=55K)& $N$(X$_i$) ($T$=132K)\\
       &            & K \kms & cm$^{-2}$         & cm$^{-2}$ \\

\hline
B0e & CS (3-2)  & 18.9 & 4.6(13) & 9.4(13) \\
B1a & CS (3-2)  & 33.1 & 8.0(13) & 1.6(14) \\
B1b & CS (3-2)  & 24.9 & 6.1(13) & 1.2(14) \\
B1c & CS (3-2)  & 28.8 & 8.0(13) & 1.4(14) \\

B0e & CH$_3$OH (3$_{1}$-2$_{1}$)E  & 1.8 & 8.8(14) & 2.2(15) \\
B1a & CH$_3$OH (3$_{1}$-2$_{1}$)E  & 3.7 & 1.9(15) & 4.7(15) \\
B1b & CH$_3$OH (3$_{1}$-2$_{1}$)E  & 6.0 & 3.0(15) & 7.4(15) \\
B1c & CH$_3$OH (3$_{1}$-2$_{1}$)E  & 2.4 & 1.2(15) & 3.0(15) \\

B0e & CH$_3$OH (3$_{0}$-2$_{0}$)A$^+$ & 11.8 & 3.6(15) & 1.1(16) \\
B1a & CH$_3$OH (3$_{0}$-2$_{0}$)A$^+$ & 36.8 & 1.1(16) & 3.6(16) \\
B1b & CH$_3$OH (3$_{0}$-2$_{0}$)A$^+$ & 48.1 & 1.5(16) & 4.6(16) \\
B1c & CH$_3$OH (3$_{0}$-2$_{0}$)A$^+$ & 33.9 & 1.0(16) & 3.3(16) \\

B0e & p-H$_2$CO (2$_{02}$-1$_{01}$) & 11.2 & 3.2(14) & 1.0(15) \\
B1a & p-H$_2$CO (2$_{02}$-1$_{01}$) & 27.7 & 8.0(14) & 2.6(15) \\
B1b & p-H$_2$CO (2$_{02}$-1$_{01}$) & 23.3 & 6.7(14) & 2.2(15) \\
B1c & p-H$_2$CO (2$_{02}$-1$_{01}$) & 16.0 & 4.6(14) & 1.2(15) \\

B0e & HC$_3$N (16-15) & 2.2 & 2.6(13) & 3.3(13) \\
B1a & HC$_3$N (16-15) & 3.6 & 4.2(13) & 5.3(13) \\
B1b & HC$_3$N (16-15) & 2.6 & 2.9(13) & 3.7(13) \\
B1c & HC$_3$N (16-15) & 2.4 & 2.7(13) & 3.5(13) \\
\hline
\end{tabular}
\label{column}
\end{table*}

\begin{table*}
\caption{Molecular column densities in the HV clumps assuming a $T$ = 64 K.}
\vspace{0.2cm}
\begin{tabular}{@{}lcccccc@{}}
\hline
clump   & $N$(HC$_3$N) & $N$(H$_2$CO) & $N$(CH$_3$OH)& $N$(CH$_3$OH) $^{\dag}$& $N$(CS) & $N$(CS)$^{\ddag}$\\ 
      & (cm$^{-2}$)  & (cm$^{-2}$) & (cm$^{-2}$) & (cm$^{-2}$)  & (cm$^{-2}$) & (cm$^{-2}$)\\
\hline
B0e--HV2 &   ...              & 1$\times$10$^{13}$ & 3$\times$10$^{14}$ &  ...  & \textgreater 4$\times$10$^{12}$ & \textgreater 4$\times$10$^{12}$ \\
B0e--HV1 & 6$\times$10$^{12}$ & 6$\times$10$^{13}$ & 1$\times$10$^{15}$ &  ...  & 3$\times$10$^{13}$ & 9$\times$10$^{13}$\\
B1a--HV  & 4$\times$10$^{12}$ & 7$\times$10$^{13}$ & 1$\times$10$^{15}$ &  ...  & 2$\times$10$^{13}$ & 2$\times$10$^{13}$\\
B1b--HV  & 1$\times$10$^{13}$ & 2$\times$10$^{14}$ & 5$\times$10$^{15}$ & 3$\times$10$^{15}$ & 2$\times$10$^{13}$ & 3$\times$10$^{13}$ \\
\hline
\end{tabular}
\label{column_bullets}
\\
$^{\dag}$ derived from rotational diagram \\
$^{\ddag}$ derived from LVG modeling
\end{table*}

\subsection{Column densities}

We estimated the column densities of the observed species toward the clumps by using the same procedure as in \citet{benedettini07} for a comparison with the 3 mm set of data.
We derived the column density of the detected species in each clump from the integrated intensity of the observed emission lines, assuming LTE condition and that the lines are optically thin. In this case the following formula can be used

\begin{equation}
N = \frac{8\times10^5 \pi k \nu^2}{h c^3 g_{up} A_{ud}} ~ Q(T_{\rm rot}) ~  {\rm exp}\left(\frac{E_{\rm up}}{kT_{\rm rot}}\right) ~ \int{T_{\rm mb} ~ dv}
\label{eqcd}
\end{equation}

where $T_{\rm rot}$ is the rotational temperature, $Q(T_{\rm rot})$ is the partition function, $g_{up}$ is the degeneracy of the upper level, $\nu$ is the frequency of the transition in GHz, $A_{ud}$ is the Einstein coefficient of the transition in s$^{-1}$, $E_{\rm up}$ is the energy of the upper level of the transition and the integral of the line emission is in K \kms. The integration limits and the polygon used to calculate the column densities are the same for all molecules in each clump. 
In Table \ref{column} we list the column densities of the observed molecules assuming the two extreme temperatures of 55 and 132 K. The two values can give an idea of the error associated to the column density produced by the uncertainty on the temperature of the gas. These values must be considered as lower limits because the LTE assumption may be not valid for most of the transitions since they have a critical density $\ge$10$^5$ cm$^{-3}$. Moreover, the total flux of the lines is underestimated due the filtering of the large scale emission (see Sect. \ref{filtering}).

Considering the large uncertainty in the column densities derived with such a procedure, that can be as high as one order of magnitude, we can say that the values derived from the present 2 mm data are consistent with the ones previously derived from 3 mm data \citep{benedettini07}, the two differing no more than a factor of 10.

We also calculated the column density of the HV bullets by using Eq. \ref{eqcd} and assuming a temperature of 64 K. The results are presented in Table \ref{column_bullets} where also the values of CS and \methanol\, column densities derived with other methods are reported (see Sect. 5.2 and 6). For each of the observed species the column densities in the four HV bullets is similar. In addition there are also no big differences between the column densities of the bullets (Table \ref{column_bullets}) and the column densities of the emission at larger scale (Table \ref{column}).

\subsection{Comparison with chemical shock models}

We compared the observed column densities of the HV bullets with chemical models of C--type shock previously used in \citet{viti11} for modeling the H$_2$O(1$_{10}$--1$_{01}$) and NH$_3$ (1$_0$--0$_0$) lines in L1157--B1. The models were run with a code that couples the UCL\_CHEM time--dependent gas--grain chemical code \citep{viti04} with the parametric shock model of \citet{jimenez08} which calculates the physical structure of a plane--parallel steady--state C--shock that propagates through an unperturbed medium. We refer to \citet{viti11} for the detailed description of the model and the choice of the investigated parameter space.
We analyzed the same parameter space as in \cite{viti11} using all the models of their Table 1. 

\begin{figure*}
\includegraphics[width=11cm, angle=90]{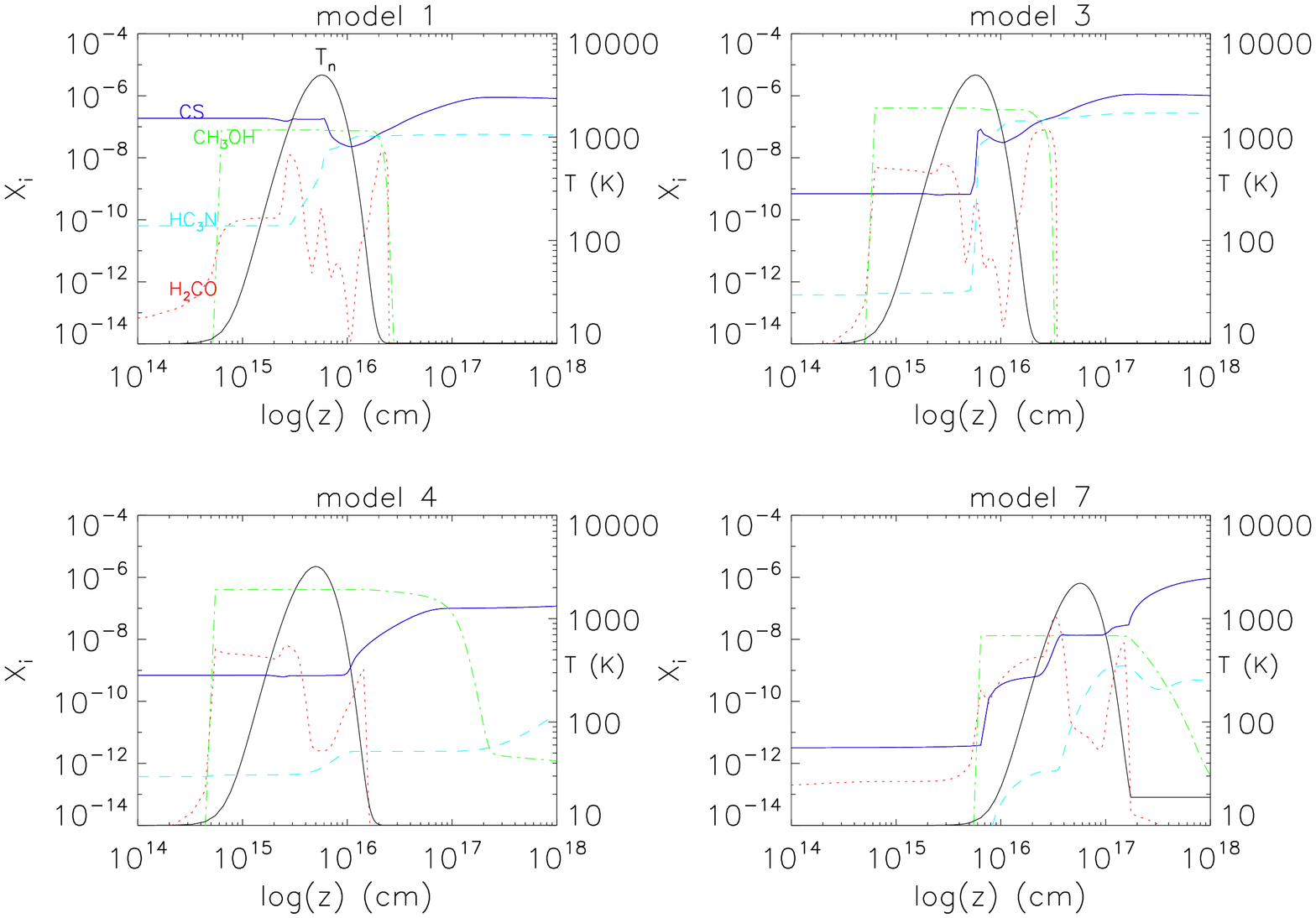}
\caption{Fractional chemical abundances with respect to H$_2$ vs the distance as predicted by the \citet{viti11} models: continuum blue for CS, dot dashed green for \methanol, dashed cyan for HC$_3$N and dotted red for H$_2$CO. The temperature profile (black) for each model is also shown.}
\label{chem_abundace}

\includegraphics[width=11cm, angle=90]{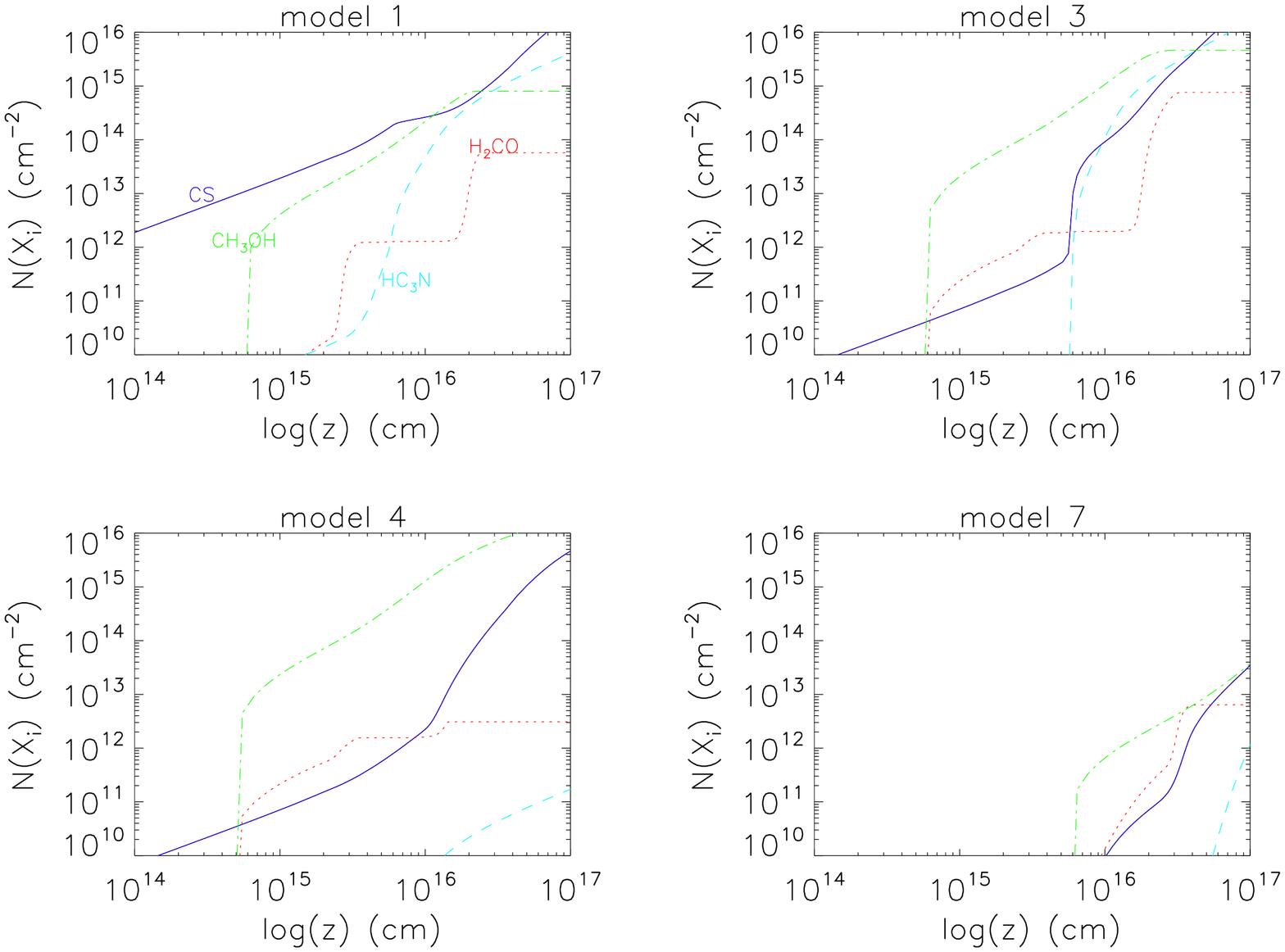}
\caption{Column densities of the observed species vs distance as predicted by the \citet{viti11} models: continuum blue for CS, dot dashed green for \methanol, dashed cyan for HC$_3$N and dotted red for H$_2$CO.}
\label{chem_cd}
\end{figure*}

In Fig. \ref{chem_abundace} we show the evolution of the chemical abundances with respect to H$_2$ as function of distance within the shock. For a direct comparison of our data with the models we calculated the theoretical column density of the observed species summing the contribution of the gas along the length of the shock (see Fig. \ref{chem_cd}). Because of the inclination of the L1157 outflow, these calculated values correspond to the column densities along the line of sight only in case of a spherical symmetry of the shocked area. Any deviation of the real shock geometry from the spherical symmetry introduces errors in the theoretical column density. However, the large uncertainties on the observed column densities, that can be as high as one order of magnitude, make useless a comparison with a more complex model of the geometry.

At the distance of L1157--B1 (250 pc, \citealt{looney07}) the average dimension of the HV clump of 5\arcsec\, corresponds to a size of 2$\times$10$^{16}$ cm while the total dimension of B1 in our maps is $\sim$20\arcsec, corresponding to about 8 $\times$10$^{16}$ cm, and it should be considered a lower limit because of the filtering of the large scale emission in interferometer (see Sect. 2.1). On the other hand, the upper limit for the size of B1 can be derived from IRAM-30m single dish observations \citep{bachiller01} to be 30\arcsec, corresponding to about 10$^{17}$ cm; hence in Fig. \ref{chem_cd} we limited the x--axis to that size. We find that models with low preshock density (10$^3$ \cmtre) have long saturation time resulting in shock length larger than 3$\times$10$^{17}$ cm which corresponds to an angular size larger than the total dimension of the blue lobe of L1157  while models with preshock density larger than 10$^4$ \cmtre\, are compatible with the dimension of the B1 shock. In particular, for models with preshock density of 10$^5$ \cmtre\, the column densities integrated over a linear dimension of $\sim$ 10$^{17}$ cm include both the thin shock layer at high temperature and the post shocked region where the temperature drops.

Models show that CH$_3$OH and H$_2$CO, species mainly formed on the grains, are sputtered back to the gas phase as soon as the shock temperature increases to a few tens of K. Therefore the abundance of these species is not directly linked to the parameters of the shock but mainly depends on the percentage of gas depleted on to the dust grains during the cold pre-shock phase. In general, models with a low percentage of depletion ($<$30\% of gas phase CO frozen onto grains at the end of the accretion phase) produce column density of CH$_3$OH and H$_2$CO a few orders of magnitude lower than observed (see for example model 1). It is interesting to note, however, that H$_2$CO has a double peak behaviour (see Fig. \ref{chem_abundace}), with the first peak occurring when the temperature starts to increases; as the shock progresses formaldehyde starts declining; it then increase again once the shock has passed through and the gas cools down. This behaviour is explained by the fact that the dominant route of destruction for this species is via a reaction with atomic hydrogen, whose abundance is directly proportional to the temperature.  

Also of particular interest is HC$_3$N, a molecule that has never been analyzed before in the framework of shock models in protostellar outflows. In fact, HC$_3$N has been observed and modeled in many pre--stellar core with fractional abundances from a few 10$^{-10}$ up to 6$\times$10$^{-8}$ in TMC-1 Core D (e.g. \citealt{ohishi98}; \citealt{tafalla06}). However, \citet{benedettini12a} showed that in pre--stellar cores HC$_3$N is quickly destroyed once the main accretion phase is finished. \citet{beltran04} observed HC$_3$N (12--11) along the wall of the cavity of the L1157 outflow close to the driving protostar and claimed that the enhancement of HC$_3$N  could be produced by the shock created by the precessing jet. Indeed, in our shock models we find that this species is formed during the shock, and only models where the gas reaches high maximum temperatures ($T_{\rm max}$=4000 K) can reproduce the observed HC$_3$N column density: HC$_3$N formation is enhanced at high temperatures mainly because of the reaction: C$_2$H + HCN which forms HC$_3$N. This reaction has a barrier of ∼ 770
K (\citealt{hoobler97}; \citealt{woodall07}). Clearly, therefore, the higher the temperature the more efficient HC$_3$N formation is. In order to reach the observed abundance, and considering the short-lived high temperature phase, models with a high T$_{\rm max}$ are favoured, in agreement with the conclusions drawn in \citet{viti11} for ammonia. It is possible of course that this species would be equally enhanced in models where a lower, but longer lived, temperature phase is maintained.

In conclusions, in agreement with \citet{viti11}, we find that the observed column densities in L1157--B1 can be reproduced qualitatively by the presence of a C--type shock with a pre--shock density $n_{\rm H_2}>$10$^4$ \cmtre\, and velocity $\sim$ 40 \kms\, so that the maximum shock temperature reaches the 4000 K.

\section{Conclusions}

We present high spatial resolution maps of the B1 shock knot in the blue lobe of the L1157 outflow of the CS (3--2), CH$_3$OH (3$_{\rm K}$--2$_{\rm K}$), HC$_3$N (16--15) and p-H$_2$CO (2$_{02}$--3$_{01}$) lines. The combined analysis of the morphology and spectral profiles has shown that the gas flowing at higher velocity is confined to a few compact ($\approx$ 5\arcsec) bullets while the gas flowing at lower velocity trace the wall of the gas cavity excavated by the jet and the apex of the B1 bow shock. 
In particular, two HV bullets, one peaking at --16 \kms\, and one at --6 \kms\,, have been detected in the direction of the north-east B0e clump, one HV bullet at --12 \kms\, has been detected in the direction of B1a and one  at --4 \kms\, in the direction of B1b.

We applied a LVG model to the CS (3--2) and (2--1) lines whose ratio is a good tracer of the gas density. By using temperature estimate from the literature we were able to give an upper limit of 10$^6$ \cmtre\, to the B1 averaged gas density, therefore refining previous density estimates that give only a lower limit of 10$^5$ \cmtre. For the compact bullets we found that the gas density is in the range 5$\times$10$^3\leqslant n_{\rm H_2} \leqslant$ 5$\times$10$^5$ \cmtre\, therefore they seem to be less dense than the large scale emitting gas. 

We derived the column densities of the observed species both for the large scale as well as for the compact gas and we found similar values. In particular, the observed column densities in L1157--B1, derived from low/medium energy transitions ($E_{\rm up} <$ 60 K) of the four molecules CS, CH$_3$OH, HC$_3$N and H$_2$CO, can be reproduced qualitatively by the presence in B1 of a C--type shock with a pre--shock density $n_{\rm H_2}>$10$^4$ \cmtre. The measured enhancement of the HC$_3$N abundance requires a maximum shock temperature of at least 4000 K that can be reached for shock velocity of $\sim$ 40 \kms.

Of particular interest is the B0e region where we found a gradient of the first moment with increasing velocity moving from the internal wall of the cavity toward the outside in the direction of B0e--HV2, the higher velocity bullet that is external with respect to the walls of the outflow cavity. The interpretation of this very intriguing feature is not univocal: it may be that the precessing jet, not yet observed directly in the L1157 outflow, is now impacting at the B0e position accelerating the gas outwards of the cavity, or, alternatively that the compact high velocity bullet is a clumpy structure formed, at least partially, before the advent of the outflow as suggested by the lower gas density in the HV bullets.

\section*{Acknowledgments}
GB and AIGR are supported by an Italian Space Agency (ASI) fellowship under contract number I/005/11/0.

\label{lastpage}

\end{document}